\documentclass[10pt,journal,compsoc]{IEEEtran}
\usepackage{CJKutf8}
\usepackage[pdfusetitle,unicode]{hyperref}
\hypersetup{unicode=true,
pdfauthor={XIE Zhibang (谢致邦) \& DENG Qingjin (邓清津)},
pdfsubject=Computer Science,
pdfkeywords={Loongson, IoT Gateway, ZigBee}}
\usepackage{tikz}
\usepackage{float}
\begin{document}
\begin{CJK}{UTF8}{gbsn}
\title{Loongson IoT Gateway: A Technical Review}
\author{XIE Zhibang (谢致邦) \& DENG Qingjin (邓清津)\\
College of Computer Science \& Software Engineering\\
Shenzhen University\\
Shenzhen, 518060, China\\
xiezhibang@email.szu.edu.cn}
\maketitle
\begin{abstract}
A prototype of Loongson IoT (Internet of Things) ZigBee gateway is already designed and implemented. However, this prototype is not perfect enough because of the lack of a number of functions. And a lot of things should be done to improve this prototype, such as adding widely used IEEE 802.11 function, using a fully open source ZigBee protocol stack to get rid of proprietary implement or using a fully open source embedded operating system to support 6LoWPAN, and implementing multiple interfaces.
\end{abstract}
\section{Introduction}
There are some kind of Loongson CPU modules that have been developed. And a prototype of Loongson IoT (Internet of Things) ZigBee gateway has been designed and implemented by ZHANG Yisu from University of Chinese Academy of Sciences, and Shenyang Institute of Computing Technology of Chinese Academy of Sciences, with other people also from Shenyang Institute of Computing Technology of Chinese Academy of Sciences \cite{zhang2013iot}. But the prototype is not perfect enough because of the lack of a number of functions.

To improve the prototype, a lot of things should be done, such as adding widely used IEEE 802.11 function (by using USB or SPI wireless NIC, or replacing Loongson 1B with Loongson 1A and using PCI wireless NIC), replacing Zstack with a fully open source ZigBee protocol stack ZBOSS to get rid of proprietary implement or using a fully open source embedded operating system Contiki to support 6LoWPAN, and implementing multiple interfaces such as Bluetooth, Infrared, and so forth.

The rest of this paper is organized as follows:
\section{Loongson}
Loongson (also called Godson) is a set of general purpose MIPS CPUs developed by the Institute of Computing Technology of Chinese Academy of Science, and Loongson Technology Corporation Limited \cite{hu2015introduction}. Loongson defined loongson 3 (Large), Loongson 2 (Medium), and Loongson 1 (Small) three series for different purpose \cite{hu2013loongson}.
\subsection{Loongson 3}
Loongson 3 (Large CPU series) is designed for servers and high performance computing applications \cite{hu2013loongson}.
\subsection{Loongson 2}
Loongson 2 (Medium CPU series) is designed for high-end embedded and computer class applications \cite{hu2013loongson}.
\subsection{Loongson 1}
Loongson 1 (Small CPU series) is designed for consumer electronics and embedded applications \cite{hu2013loongson}. Loongson 1 can be supported by Linux, VxWorks, RT-Thread, and other operating systems \cite{hu2003ls1,yuan2009radiation,xie2010loongson}.
\subsubsection{Loongson 1A}
Loongson 1A (also called LS1A) is using 0.13um technology, and integrated a LS232 processor core, 2D GPU, 16/32-bit DDR2, High-definition display, PCI, Southbridge chipset features, RSECC NAND, CAN, ACPI, SPI, 88-way GPIO interfaces, and so forth. Loongson 1A supports LPC/SPI/NAND start-up mode. The high level integration of Loongson 1A makes Loongson 1A particularly suitable for cloud terminals, industrial control, data acquisition, network device and other fields. Meanwhile Loongson 1A can be configured to a full-featured Southbridge chip with PCI interface \cite{hu2013loongson}.
\subsubsection{Loongson 1B}
Loongson 1B (also called LS1B) is using 0.13um technology, and it is a lightweight 32 SoC chip. It integrated a LS232 processor core, 16/32-bit DDR2, high-definition display, NAND, SPI, 62-way GPIO, USB, CAN, UART interfaces, and so forth. And it is able to meet the needs of low price cloud terminals, data collection, network device and other fields \cite{hu2013loongson}.
\subsubsection{Loongson 1C}
Loongson 1C (also called LS1C) is using 0.13um technology. Loongson 1C not only integrated a 32-bit Loongson processor core, but also integrated a 32/16/8 bit SDRAM, LCD display, CAMERA, USBHOST, USB-OTG, SDIO, ADC, RSECCNAND, CAN, I2C, I2S/AC97, SPI, UART, 102 road GPIO interfaces, and so forth. Loongson 1C supports SPI/NAND/SDIO start-up mode \cite{hu2013loongson}.
\subsubsection{Loongson 1D}
Loongson 1D (also called LS1D) is using 0.13um technology, and integrated a 32-bit Loongson processor core, 64KB Flash, 5KB RAM, Chopper stabilized amplifiers, High-precision time-digital converter (TDC), SPI, UART, I2C and other functions. Loongson 1D is a single-chip solution for water meters, gas meters and other flow meters, with high precision and low power consumption \cite{hu2013loongson}.
\subsubsection{Loongson 1E}
Loongson 1E (also called LS1E) is designed for aerospace applications, and used radiation hardening technology. It is according to the previous generations to customized for aerospace applications from system design \cite{fang2013space}.
\subsubsection{Loongson 1F}
Loongson 1F (also called LS1F) is the companion IO bridge chip for Loongson 1E. It integrated commonly used interfaces in aerospace field, including remote sensing and control functions interfaces and peripheral interfaces, to companion Loongson 1E \cite{fang2013space}.
\section{Loongson IoT Gateway}
ZHANG Yisu from University of Chinese Academy of Sciences, and Shenyang Institute of Computing Technology of Chinese Academy of Sciences, with other people also from Shenyang Institute of Computing Technology of Chinese Academy of Sciences, have already designed and implemented a prototype of Loongson IoT ZigBee Gateway. The prototype based on Loongson 1B develop board, TI CC2530 ZigBee develop kit, Zstack protocol stack, and embedded Linux system \cite{zhang2013iot}.
\subsection{Hardware structure}
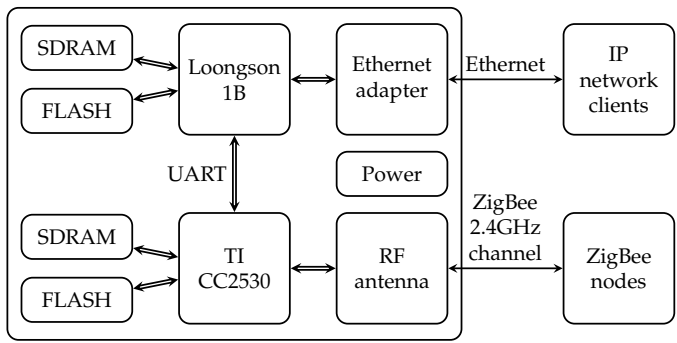
\begin{figure}[H]
\resizebox{\linewidth}{!}
{\begin{tikzpicture}
[c/.style={align=center},
f/.style={draw,thick,rounded corners=5},
r/.style={f,minimum width=50},
r1/.style={r,minimum height=20},
r2/.style={r,minimum height=50,c},
s/.style={stealth-stealth,thick},
sd/.style={s,double}]
\node[f,minimum width=205,minimum height=150]at(3.5,2.5){};
\node[r1](tf)at(1,0.5){FLASH};
\node[r1](ts)at(1,1.5){SDRAM};
\node[r2](ti)at(3.5,1){TI\\CC2530};
\draw[sd](tf)--(ti);
\draw[sd](ts)--(ti);
\node[r1](lf)at(1,3.5){FLASH};
\node[r1](ls)at(1,4.5){SDRAM};
\node[r2](l1)at(3.5,4){Loongson\\1B};
\draw[sd](ti)--node[left]{UART}(l1);
\draw[sd](lf)--(l1);
\draw[sd](ls)--(l1);
\node[r2](ea)at(6,4){Ethernet\\adapter};
\draw[sd](ea)--(l1);
\node[r2](ra)at(6,1){RF\\antenna};
\draw[sd](ra)--(ti);
\node[r1]at(6,2.5){Power};
\node[r2](pc)at(9.6,4){IP\\network\\clients};
\draw[s](ea)--node[above]{Ethernet}(pc);
\node[r2](zn)at(9.6,1){ZigBee\\nodes};
\draw[s](ra)--node[above,c]{ZigBee\\2.4GHz\\channel}(zn);
\end{tikzpicture}}
\caption{The hardware structure of the prototype.}
\label{fig:hardware}
\end{figure}

The hardware structure of the prototype is shown in the Figure \ref{fig:hardware}.

TI CC2530 connects to Loongson 1B using UART interface, and Loongson 1B communicates to PC through Ethernet with Ethernet adapter, and TI CC2530 communicates to ZigBee nodes through ZigBee 2.4 GHz channel with RF antenna \cite{zhang2013iot}.
\subsection{Software structure}
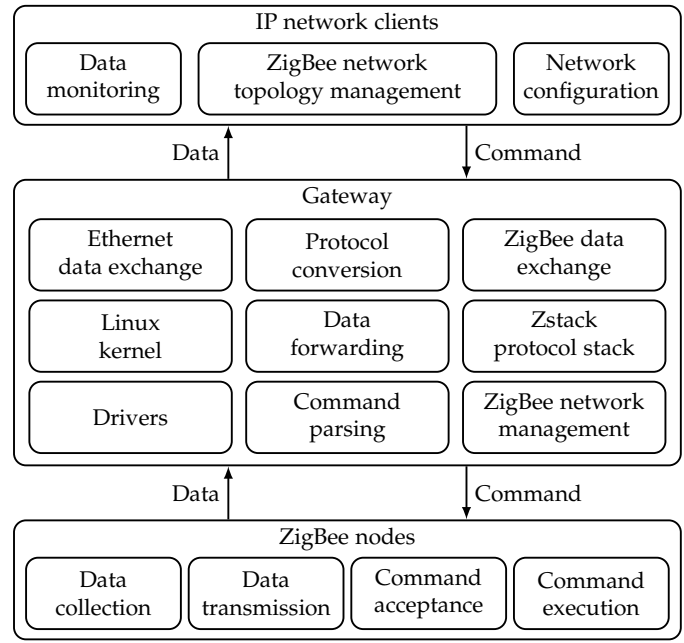
\begin{figure}[H]
\resizebox{\linewidth}{!}
{\begin{tikzpicture}
[c/.style={align=center},
r/.style={draw,thick,c,rounded corners=5},
n/.style={r,minimum height=30},
n1/.style={n,minimum width=65},
n2/.style={n,minimum width=85},
n3/.style={n,minimum width=125},
f/.style={r,minimum width=280},
f1/.style={f,minimum height=50},
f2/.style={f,minimum height=120},
s/.style={-latex,thick}]
\node[f1](zn)at(5,1){};
\node[below]at(zn.north){ZigBee nodes};
\node[n1]at(1.4,0.8){Data\\collection};
\node[n1]at(3.8,0.8){Data\\transmission};
\node[n1]at(6.2,0.8){Command\\acceptance};
\node[n1]at(8.6,0.8){Command\\execution};
\node[f2](gw)at(5,4.8){};
\node[below]at(gw.north){Gateway};
\draw[s]([xshift=-50]zn.north)--node[left]{Data}([xshift=-50]gw.south);
\draw[s]([xshift=50]gw.south)--node[right]{Command}([xshift=50]zn.north);
\node[n2]at(1.8,3.4){Drivers};
\node[n2]at(5,3.4){Command\\parsing};
\node[n2]at(8.2,3.4){ZigBee network\\management};
\node[n2]at(1.8,4.6){Linux\\kernel};
\node[n2]at(5,4.6){Data\\forwarding};
\node[n2]at(8.2,4.6){Zstack\\protocol stack};
\node[n2]at(1.8,5.8){Ethernet\\data exchange};
\node[n2]at(5,5.8){Protocol\\conversion};
\node[n2]at(8.2,5.8){ZigBee data\\exchange};
\node[f1](nc)at(5,8.6){};
\node[below]at(nc.north){IP network clients};
\draw[s]([xshift=-50]gw.north)--node[left]{Data}([xshift=-50]nc.south);
\draw[s]([xshift=50]nc.south)--node[right]{Command}([xshift=50]gw.north);
\node[n1]at(1.4,8.4){Data\\monitoring};
\node[n3]at(5,8.4){ZigBee network\\topology management};
\node[n1]at(8.6,8.4){Network\\configuration};
\end{tikzpicture}}
\caption{The software structure of the prototype.}
\label{fig:software}
\end{figure}

The software structure of the prototype is shown in the Figure \ref{fig:software}.

The software of the prototype is based on embedded Linux operating system and Zstack protocol stack \cite{zhang2013iot}.
\section{Analysis}
The prototype of Loongson IoT ZigBee Gateway already has the basic function of IoT Gateway, but it still has a lot of things that can be improved:
\subsection{IEEE 802.11}
IEEE 802.11 (also called Wi-Fi) is widely used. And as an IoT Gateway, IEEE 802.11 should be one of the basic functions. But the prototype lacks it.

Because of Loongson 1B lacking PCI interface \cite{hu2013loongson}, Loongson 1B just can use USB wireless NIC to support high speed IEEE 802.11 network up to 480 Mbps and use SPI low energy wireless NIC to support low speed IEEE 802.11 network up to 30 Mbps.

To support IEEE 802.11 network upper than 480 Mbps, the gateway should use Loongson 1A instead of Loongson 1B to use PCI wireless NIC \cite{hu2013loongson}. When the PCI interface of Loongson 1A running in 32-bit at 33 MHz, the speed up to 1064 Mbps, and when the PCI interface of Loongson 1A running in 32-bit at 66 MHz, the speed up to 2128 Mbps.
\subsection{Fully open source ZigBee protocol stack}
The prototype is using Zstack as its protocol stack, but Zstack is not fully open source, the implement of MAC layer and network layer is provided as precompiled library files \cite{diao2012osgi}.

To get rid of proprietary implement of MAC layer and network layer, ZBOSS (ZigBee Open Source Stack) v1.0 is a nice choice. ZBOSS v1.0 is fully open source \cite{ma2014zigbee} and it supports TI CC253x series chips.
\subsection{6LoWPAN and Fully open source embedded OS}
6LoWPAN, IPv6 networking over low rate personal area networks, will be the future of IoT, because of the address space exhaustion of IPv4 \cite{shelby20116lowpan}.
Contiki, a fully open source embedded operating system using uIP TCP/IP (v4 and v6) stack and Rime stack supports 6LoWPAN \cite{dunkels2004contiki,yang2013contiki}.
TI CC2530 can use Contiki to support 6LoWPAN \cite{yang2013contiki}. 
\subsection{Multiple interfaces}
A lot of IoT devices are using specific network communication standards, such as IEEE 802.11, ZigBee, Bluetooth, Infrared, and so forth. To solve this problem, the Loongson IoT gateway should support multiple interfaces \cite{chang2015iot}.
Because of Loongson 1B having 12 UART interfaces and 4 PWM interfaces \cite{lpwp}, it can connect with lots of UART and PWM chips, such as a UART ZigBee chip, a UART Bluetooth chip, a PWM Infrared chip, and so forth.

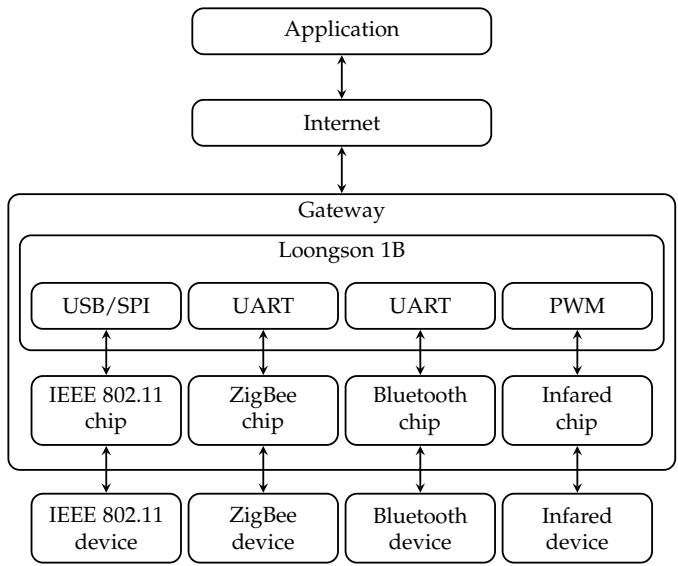
\begin{figure}[H]
\resizebox{\linewidth}{!}
{\begin{tikzpicture}
[r/.style={draw,thick,align=center,rounded corners=5},
r1/.style={r,minimum width=65},
r2/.style={r1,minimum height=30},
r3/.style={r1,minimum height=20},
r4/.style={r,minimum width=130,minimum height=20},
s/.style={stealth-stealth,thick}]
\node[r2](8d)at(0,0){IEEE 802.11\\device};
\node[r2](8c)at(0,1.8){IEEE 802.11\\chip};
\draw[s](8d)--(8c);
\node[r3](us)at(0,3.4){USB/SPI};
\draw[s](us)--(8c);
\node[r2](zd)at(2.4,0){ZigBee\\device};
\node[r2](zc)at(2.4,1.8){ZigBee\\chip};
\draw[s](zd)--(zc);
\node[r3](u1)at(2.4,3.4){UART};
\draw[s](u1)--(zc);
\node[r2](bd)at(4.8,0){Bluetooth\\device};
\node[r2](bc)at(4.8,1.8){Bluetooth\\chip};
\draw[s](bd)--(bc);
\node[r3](u2)at(4.8,3.4){UART};
\draw[s](u2)--(bc);
\node[r2](id)at(7.2,0){Infared\\device};
\node[r2](ic)at(7.2,1.8){Infared\\chip};
\draw[s](id)--(ic);
\node[r3](p)at(7.2,3.4){PWM};
\draw[s](p)--(ic);
\node[r,minimum width=280,minimum height=50](ls)at(3.6,3.6){};
\node[below]at(ls.north){Loongson 1B};
\node[r,minimum width=290,minimum height=120](gw)at(3.6,3){};
\node[below]at(gw.north){Gateway};
\node[r4](in)at(3.6,6.2){Internet};
\draw[s](gw)--(in);
\node[r4](ap)at(3.6,7.6){Application};
\draw[s](ap)--(in);
\end{tikzpicture}}
\caption{The multiple interfaces structure of the Loongson 1B IoT gateway.}
\label{fig:ls1b}
\end{figure}

The multiple interfaces structure of the Loongson 1B IoT gateway is shown in the Figure \ref{fig:ls1b}.

However, Loongson 1B with SPI wireless NIC only supports IEEE 802.11 network up to 30 Mbps, and with USB wireless NIC only supports IEEE 802.11 network up to 480 Mbps. To support higher speed IEEE 802.11 network, the Loongson IoT gateway should use Loongson 1A instead of Loongson 1B, with a PCI wireless NIC. Because of Loongson 1A having 4 UART interfaces and 4 PWM interfaces \cite{lpwp}, it can also connect with lots of UART and PWM chips.

\begin{figure}[H]
\resizebox{\linewidth}{!}
{\begin{tikzpicture}
[r/.style={draw,thick,align=center,rounded corners=5},
r1/.style={r,minimum width=65},
r2/.style={r1,minimum height=30},
r3/.style={r1,minimum height=20},
r4/.style={r,minimum width=130,minimum height=20},
s/.style={stealth-stealth,thick}]
\node[r2](8d)at(0,0){IEEE 802.11\\device};
\node[r2](8c)at(0,1.8){IEEE 802.11\\chip};
\draw[s](8d)--(8c);
\node[r3](us)at(0,3.4){PCI};
\draw[s](us)--(8c);
\node[r2](zd)at(2.4,0){ZigBee\\device};
\node[r2](zc)at(2.4,1.8){ZigBee\\chip};
\draw[s](zd)--(zc);
\node[r3](u1)at(2.4,3.4){UART};
\draw[s](u1)--(zc);
\node[r2](bd)at(4.8,0){Bluetooth\\device};
\node[r2](bc)at(4.8,1.8){Bluetooth\\chip};
\draw[s](bd)--(bc);
\node[r3](u2)at(4.8,3.4){UART};
\draw[s](u2)--(bc);
\node[r2](id)at(7.2,0){Infared\\device};
\node[r2](ic)at(7.2,1.8){Infared\\chip};
\draw[s](id)--(ic);
\node[r3](p)at(7.2,3.4){PWM};
\draw[s](p)--(ic);
\node[r,minimum width=280,minimum height=50](ls)at(3.6,3.6){};
\node[below]at(ls.north){Loongson 1A};
\node[r,minimum width=290,minimum height=120](gw)at(3.6,3){};
\node[below]at(gw.north){Gateway};
\node[r4](in)at(3.6,6.2){Internet};
\draw[s](gw)--(in);
\node[r4](ap)at(3.6,7.6){Application};
\draw[s](ap)--(in);
\end{tikzpicture}}
\caption{The multiple interfaces structure of the Loongson 1A IoT gateway.}
\label{fig:ls1a}
\end{figure}
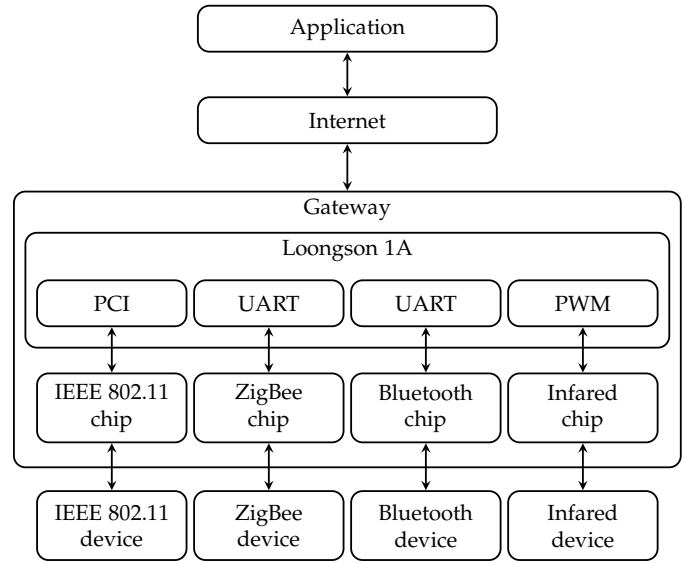

The multiple interfaces structure of the Loongson 1A IoT gateway is shown in the Figure \ref{fig:ls1a}.

\section{Conclusions}
Although the prototype of Loongson IoT ZigBee Gateway already has the basic function of IoT Gateway, it lacks a number of functions, and a lot of things still can be done to improve it, such as adding widely used IEEE 802.11 function (by using USB or SPI wireless NIC, or replacing Loongson 1B with Loongson 1A and using PCI wireless NIC), replacing Zstack with a fully open source ZigBee protocol stack ZBOSS to get rid of proprietary implement or using a fully open source embedded operating system Contiki to support 6LoWPAN, and implementing multiple interfaces such as Bluetooth, Infrared, and so forth.
\section{Acknowledgement}
The authors wish to acknowledge the help of their Graduate Teaching Assistant, Mr. Salman Salloum in commenting on an early draft of the chapter.
\bibliography{loongson_iot_gateway}
\bibliographystyle{IEEEtran}
\end{CJK}
\end{document}